\documentclass{article}         
\begin{document}
\title{Nonlinear and Quantum Origin of Doubly Infinite Family of 
Modified Addition Laws for 
Fourmomenta
\thanks{Supported by KBN grant 5PO3B05620}}
\author{J. Lukierski
\\
Institute for Theoretical Physics, University of Wroc{\l}aw,
 \\
 pl. Maxa Borna 9, 50--205 Wroc{\l}aw, Poland
 \\ \\
A. Nowicki
\\
Institute of Physics, University of Zielona G\'{o}ra,
 \\
 pl. S\l owia\'{n}ski 6, 65-069 Zielona G\'{o}ra, Poland}
\maketitle
\begin{abstract}
We show that infinite variety of Poincar\'{e} bialgebras with
nontrivial classical r-matrices generate nonsymmetric nonlinear
composition laws for the fourmomenta. We also present the 
problem
of lifting the Poincar\'{e} bialgebras to quantum Poincar\'{e}
groups by using e.g. Drinfeld twist, what permits to provide the
nonlinear composition law in any order of dimensionfull
deformation parmeter $\lambda$ (from physical reasons we can put
$\lambda = \lambda _{p}$ where $\lambda_{p}$ is the Planck
lenght). The second infinite variety of composition laws for
fourmomentum is obtained by nonlinear change of basis in
Poincar\'{e} algebra, which can be performed for any choice of
coalgebraic sector, with classical or quantum coproduct. In last
Section we propose some modification of Hopf algebra scheme with
Casimir-dependent deformation parameter, which can help to 
resolve
the problem of consistent passage to macroscopic classical limit.

\end{abstract}
\section{Introduction}

\renewcommand{\theequation}{\thesection.\arabic{equation}}

The standard generators of classical relativistic symmetries are
described by classical Poincar\'{e} algebra\footnote{We shall
consider further the case D=4, but all arguments in principle can
be extended to any D } ($\eta_{\mu\nu} = (1,-1,-1,-1)$)

\begin{eqnarray}
[M^{(0)}_{\mu\nu}, M^{(0)}_{\rho\tau}] & = &  i(\eta_{\mu
\tau}M^{(0)}_{\nu\rho} - \eta_{\nu\tau}M^{(0)}_{\mu\rho} \cr\cr &&
+ \ \eta_{\nu\rho}M^{(0)}_{\mu\tau} -
\eta_{\kappa\rho}M^{(0)}_{\nu\tau})\, , \cr\cr [M^{(0)}_{\mu\nu},
P^{(0)}_{\rho}] & = & i ( \eta_{\nu\rho}P^{(0)}_{\kappa} -
\eta_{\mu\rho}P^{(0)}_{\nu})\, , \cr\cr [P^{(0)}_{\kappa},
P^{(0)}_{\nu}]& = & 0 \, .
\label{lupra1}
\end{eqnarray}
where fourmomenta generators $P^{(0)}_{\nu}=
(\overrightarrow{P}^{(0)},E^{(0)} = c P^{(0)}_0 ) $ provide the
physical three-momentum as well as energy operators, and Lorentz
generators $M^{(0)}_{\mu\nu}= ( \overrightarrow{M},
\overrightarrow{N}) $describe relativistic angular momentum. The
classical Poincar\'{e} symmetries are described by a bialgebra
with the relations (\ref{lupra1})
 satisfied by the primitive coproduct $(I^{(0)}_A = (P^{(0)}_\mu,
 M^{(0)}_{\mu\nu})$)
\begin{equation}\label{lupra2}
  \Delta(I^{(0)}_A )= I^{(0)}_A \otimes 1 + 1 \otimes I^{(0)}_A\,
  .
\end{equation}
The coproduct (\ref{lupra1}) describes classical Abelian addition
law of the  fourmomenta  and the  relativistic angular momenta.
For the system described by tensor product $ |1> \otimes |2>$ of
 the irreducible representations $|i> $ $(i=1,2)$ of the algebra
(\ref{lupra1}) one gets
\begin{eqnarray}\label{lupra3a}
  p^{(0)}_{\mu;1+2} & = & \ p^{(0)}_{\mu;1} + \ p^{(0)}_{\mu;2}\,
  ,\\ \label{lupra3b}
M^{(0)}_{\mu\nu}  & =  &\  M^{(0)}_{\mu\nu;1}  + \
M^{(0)}_{\mu\nu;2}\, . \label{lupra3}
\end{eqnarray}
where $P^{(0)}_\mu |i> = p^{(0)}_{\mu;i} |i>$ and $ P_\mu
(|1>\otimes |2>) = p^{(0)}_{\mu;1+2}(|1>\otimes |2>)$ etc.

If the tensor product $|1>\otimes |2>$ describe a composite
system,  from (\ref{lupra3a}--\ref{lupra3b}) follows that both its
constituents $|1>$ and $|2>$ are not interacting with each other.
It appears however interesting to look for modification of
classical relativistic symmetry scheme in such a way that the
derived composition law will produce at least in the relation
(\ref{lupra3a}) some primary geometric interaction term
\begin{equation}\label{lupra4}
  p_{\mu;1+2} = p_{\mu;1} + p_{\mu;2} +
  p_{\mu;1+2}(p_{\mu;1},p_{\mu,2})\, .
\end{equation}
If we assume that the origin of this geometric interaction is due
to the replacement of classical Minkowski geometry by quantum
geometry generated by quantum gravity effects it is plausible to
assume that the ``corrections''  to the formula (\ref{lupra3a})
are of order  $\frac{1}{M_{p}}$ (the inverse of Planck mass).

There are two ways of modifying the standard framework of
description of fourmomenta in classical relativistic theory,
which leads to the modification (\ref{lupra4}):

i) One can assume that the physical generators $P_\mu$ 
describing
energy and three--momenta are nonlinear functions of the classical
generators $P^{(0)}_\mu$\footnote{We shall restrict our
considerations to the case when the Lorentz sector of
(\ref{lupra1}) is not modified.}. The additional requirement of
D=3 nonrelativistic classical covariance ($M_i = \frac{1}{2}
\varepsilon_{ijk} M_{jk})$
\begin{equation}\label{lupra5}
  [M_i , M_j ] = i \varepsilon_{ijk}M_k \, , \qquad \qquad
  [M_i , P_j ] = i\varepsilon_{ijk}P_k \, ,
\end{equation}
implies that
\begin{equation}\label{lupra6}
P_\mu = P_\mu ( \overrightarrow{P}^{(0)2} , E^{(0)})\, ,\qquad
\qquad M_{\mu\nu} = M^{(0)}_{\mu\nu}\, .
\end{equation}
Such nonlinear transformations, considered recently in
\cite{ms}--\cite{ag} shall be discussed in Sect. 2. We would like
to stress here that in such an approach the symmetries remain
classical i.e. they are described by the  same classical
Poincar\'{e} bialgebra (\ref{lupra1}--\ref{lupra2}). The formulae
(\ref{lupra6}) one gets if we choose different basis  in
enveloping classical Poincar\'{e} bialgebra.

ii) Second way consists in introducing nonclassical Poincar\'{e}
bialgebras differing from (\ref{lupra1}) in coalgebra sector, with
coproduct determined by classical Poincar\'{e} r-matrces. These
matrices were almost completely classified by S. Zakrzewski
\cite{zak}. In Sect. 3 we shall select the infinite variety of
them, which can be used for the consistent deformation of the
addition law (\ref{lupra3a}).

 It appears that the knowledge of classical
r-matrix only permits to determine the modified addition law in
lowest order in $ \frac{1}{M_p}$. In order to    obtain  the
addition law for fourmomenta valid in any order in the powers of
 the deformation parameter (chosen here to be $ \frac{1}{M_p}$, 
with
the classical ``no deformation'' limit given by $M_p \to \infty$)
we should extend the quantum bialgebra to Hopf
algebra\footnote{Sometimes this extension is called ''quantization
of bialgebra'' (see e.g. \cite{ek}).}.

If we keep the classical Poincar\'{e} algebra sector unchanged,
the Poincar\'{e} r-matrices which are solutions of classical
Yang-Baxter equation (CYBE) imply the Hopf-algebraic structure in
coalgebraic sector described by the twist function, firstly
introduced by Drinfeld (see e.g. \cite{dr2}--\cite{klm}).
Unfortunately we do not know explicitly all Hopf-algebraic
counterparts of known Poincar\'{e} bialgebras listed in
\cite{zak}. We recall here that the example for which the complete
Hopf algebra structure is known in closed form and applied to
deformations of relativistic symmetries is the so--called
$\kappa$-deformed  quantum Poincar\'{e} algebra (see e.g.
\cite{lnt}--\cite{lrz}) and its generalizations \cite{km},
\cite{llm}\footnote{We would like to  point out here that several
authors restricted possible deformations of addition law for
relativistic fourmomenta only to the ones coming from classical
bialgebra (\ref{lupra1}--\ref{lupra2}) and $\kappa$-deformed
Poincar\'{e} algebra. In principle there is however  infinite
variety of such deformations.}.

One can say that the Drinfeld twists which do not commute with the
classical coproducts (\ref{lupra2}) for $P_\mu$ introduce
different nonlinear extensions of the addition law (\ref{lupra3a})
for the fourmomenta. It should be stressed however, that to the
variety of quantum deformations one can supplement the variety of
nonlinear transformations of basic generators. In this paper we
wish to point out clearly this two basic origins of possible
deformed addition laws. Finally we shall also comment on recent
efforts to describe the  addition formulae for fourmomenta outside
of Hopf-algebraic scheme which can be defined as proposals of
representation-dependent Hopf algebra structure. It appears that
in such a scheme one can look for the solution of the problem of
consistent transition from quantum-deformed Planck scale regime 
to
classical relativistic systems at macroscopic distances.

\section{Nonlinear Fourmomentum Basis}
\setcounter{equation}{0}

Let us consider firstly the classical    Poincar\'{e}  bialgebra
described by (\ref{lupra1})\footnote{In fact the classical
Poincar\'{e}  bialgebra is also a Hopf algebra, with antipode
(coinverse) given by formula $S(I_A) = I_A$.}. We shall assume
that the transformation (\ref{lupra6}) has its inverse
\begin{equation}\label{lupra2.1}
  P^{(0)}_\mu = P^{(0)}_\mu ( \overrightarrow{P}^2 , E )\, .
\end{equation}
Having formulae (\ref{lupra6}) and (\ref{lupra2.1}) one can
calculate the coproduct of $P_\mu$. Taking into consideration that
in the framework of bialgebras and Hopf algebras the coproduct is
homomorphic, i.e. $\Delta(f(I_A))= f(\Delta (I_A))$\footnote{The
proof is demonstrated rigorously for holomorphic functions $f$ and
with some requirements on the domain of generators $I_A$.}, one
gets
\begin{equation}\label{lupra2.2}
  \Delta( P_\mu ) = P_\mu    (P^{(0)}_\mu (P)\otimes 1 + 1 \otimes
  P^{(0)}_\mu (P))\, .
\end{equation}
and we obtain the following addition law for the classical
fourmomenta in nonlinear basis (\ref{lupra6})
\begin{equation}\label{lupra2.3}
  P_{\mu;1+2} = P_\mu (P^{(0)}_\mu (P_{\mu;1}) + P^{(0)}_\mu
  (P_{\mu;2}))\, .
\end{equation}
The rule (\ref{lupra2.3}) has been firstly demonstrated for
nonlinear bases of classical Poincar\'{e} algebra in  \cite{ln2},
and further rediscovered in \cite{jv}. The characteristic property
of the composition law (\ref{lupra2.3}) is its symmetry, what
discloses the classical nature of symmetries. It should by also
stressed that due to the  coassociativity  of the coproduct
\begin{equation}\label{lupra2.4}
  (\Delta \otimes 1) \Delta(P_{\mu}) = \Delta \cdot (\Delta
  \otimes 1)(P_\mu )\, ,
\end{equation}
the addition law (\ref{lupra2.3}) is coassociative, i.e. if we
denote \cite{ss1}
\begin{eqnarray*}
  \Delta(P_\mu) = \sum\limits_{i} f^i (P_\mu)  \otimes g^i 
(P_\mu)\Longleftrightarrow
  P_{\mu;1 +2} = \sum\limits_{i} f^{i} (P_{\mu;1}) \cdot g^i
  (P_{\mu;2})\, ,
\end{eqnarray*}
the coassociativity implies that for any nonlinear basis we get
\begin{equation}\label{lupra2.5}
  P_{\mu;1+2+3}= P_{\mu;1+(2+3)} = P_{\mu;(1+2)+3}\, ,
\end{equation}
where
\begin{eqnarray}\label{lupra2.6}
  P_{\mu;1+(2+3)} = \sum\limits_{i} f^i (P_{\mu;1}) \cdot g^i
  (P_{\mu;2+3})\, ,
  \cr
   P_{\mu;(1+2)+3} = \sum\limits_{i} f^i (P_{\mu;1+2}) \cdot g^i
  (P_{\mu;3})\, .
\end{eqnarray}

One of the consequences of the nonlinear change (\ref{lupra2.1})
of fourmomentum basis is the modification of mass Casimir
\begin{equation}\label{lupra2.7}
  C_2 = P^{(0)}_\mu P^{(0)\mu} \to C_2  = P^{(0)}_\mu (P)
  P^{(0)\mu}(P)\, .
\end{equation}
In physical applications one usually assumes that
\begin{equation}\label{lupra2.8}
  P_\mu  (P^{(0)}_\mu) = P^{(0)}_\mu + \frac{1}{M_p} f^{(1)}_\mu
  (P^{(0)}_\mu)  + {\cal O} ( \frac{1}{M_{p}^2})
\end{equation}
implying

\begin{equation}\label{lupra2.9}
  P^{(0)}_\mu  (P_\mu) = P_\mu - \frac{1}{M_p} f^{(1)}_\mu
  (P_\mu)  + {\cal O} ( \frac{1}{M_{p}^2})\, ,
\end{equation}
where $(f^{(1)}_\mu)^{-1} f^{(1)}_\nu = \delta_{\mu\nu} +{\cal O}
( \frac{1}{M_{p}^2})$. A known example of the transformations
(\ref{lupra6}) and (\ref{lupra2.1}) relating the standard
mass-shell condition with $\kappa$-deformed mass-shells
\cite{lrz}\footnote{We  put following further applications $\kappa
= M_p$} (we assume the light velocity $c=1$)
\begin{equation}\label{lupra2.10}
  C_2 = \left(  2M_p \sinh  \frac{E}{2M_p}\right)^2 -
  \overrightarrow{p}^2 \, e^{\pm \frac{E}{M_p}}\, ,
\end{equation}
are given in \cite{klms}.

It should be stressed that the change of nonlinear  basis,
defining the explicite form of modified mass shell condition (see
(\ref{lupra2.7})) does not imply the form of the  coproduct, which
depends on the choice of coalgebraic sector, determined by
classical r-matrix. In particular doubly   special relativistic
theories proposed in  [18,19] for any choice of deformed mass
shell condition  can be endowed by infinite variety of
fourmomentum addition laws, parametrized in lowest order  of the
deformation parameter by Poincar\'{e} classical r-matrices
\cite{zak}.

\section{Poincar\'{e} Bialgebras, Poincar\'{e} Hopf Algebras and
Addition Law for the Fourmomenta}

\setcounter{equation}{0}

A second cause which provides the modification of addition law for
the fourmomenta is due to different possible choices of coalgebra
sector for the Poincar\'{e} bialgebras. The class of Poincar\'{e}
r-matrices which describe the deformation with dimensionfull
parameters has the following general form\footnote{The classical
r-matrix $r=r^{\mu;\nu} P^{(0)}_\mu \wedge P^{(0)}_\nu$ describes
so-called soft deformation \cite{zak2} with   trivial CYBE and
will be not considered here.}

\begin{equation}\label{lupra3.1}
{\bf r}_{(i)} = \, i\, r^{\mu;\rho\tau}_{(i)}  P^{(0)}_{\mu}\wedge
M^{(0)}_{\rho\tau}\, .
\end{equation}
and satisfying  the Yang--Baxter equation (see \cite{zak}). Let us
observe that in order to obtain the   dimensionless quantity we
should multiply (\ref{lupra3.1}) by an inverse of mass parameter
$\chi$, which may be  put equal to $\frac{1}{M_p}$.

In particular we obtain the following modification of the
coproduct (\ref{lupra1})

\begin{equation}\label{lupra3.2}
  \Delta^{\kappa}_{(i)} (I_A) =
  I_A \otimes 1 + 1 \otimes I_A  +  \frac{1}{\kappa}
[ I_A \otimes 1 + 1 \otimes I_A, {\bf r}_{(i)} ]\, .
\end{equation}
where lower index $(i)$ enumerates different choices of 
 classical r-matrices.
In particular we obtain
\begin{equation}\label{lupra3.3}
  \Delta^{\kappa}_{(i)} (P_\mu) =
  P_\mu \otimes 1 + 1 \otimes P_\mu + \frac{2}{\kappa}
  r^{\mu;\rho \nu}_{(i)}
  (P_\nu \otimes P_\rho - P_\rho \otimes P_\nu )\, ,
\end{equation}
implying the following addition rule
\begin{equation}\label{lupra3.4}
p_{\mu;1+2} = p_{\mu;1}  + p_{\mu;2}  + \frac{2}{\kappa}
r^{\mu;\rho\nu}_{(i)} (p_{\nu;1}  p_{\rho;2} - p_{\rho;1}
p_{\nu;2} )\, .
\end{equation}

All these modification addition laws are nonsymmetric. Among the
variety of classical r -matrices in \cite{zak} there is one
determining the $ \kappa$-deformation, given by the formula ($N_i
\equiv M_{i0}$) (see e.g. \cite{mr1})
\begin{equation}\label{lupra3.5}
{\bf r}_{(\kappa)} = \frac{1}{\kappa}\, N_{i} \wedge P_i \, .
\end{equation}
It provides the following fourmomentum composition law
\begin{equation}\label{lupra3.6}
E_{1+2} = E_{1} + E_{2} \, , \quad \overrightarrow{p}_{1+2} =
\overrightarrow{p}_{1} +  \overrightarrow{p}_{2} +
\frac{1}{\kappa} ( E_{1} \overrightarrow{p}_{2} - E_{2}
\overrightarrow{p}_{1} )\, .
\end{equation}
The formulae (\ref{lupra3.4}) and (\ref{lupra3.6}) describe the
modified coproduct as well as the addition law for the fourmomenta
only in first order of the deformation parameter\footnote{One can
say that bialgebras describe infinitesimal form of quantum
group.}. The complete deformation formula is provided by promoting
bialgebra to quantum algebra -- a noncocommutative Hopf algebra
(see e.g. \cite{smaj}). In such a case  the quantum nonsymmetric
coproduct satisfies the relations
\begin{equation}\label{lupra3.7}
\tau (\Delta (I_A)) = \widehat{R}^{-1} \otimes \Delta (I_A)
\otimes \widehat{R}\, .
\end{equation}
where $\tau$ denotes flip operator ($\tau (A\otimes B) = B\otimes
A$) and $ \widehat{R}$ is the quantum r-matrix (called also
universal R-matrix), which satifies quantum Yang-Baxter equation
and in first order in $1/\kappa$ is desribed by a classical
r-matrix:
\begin{equation}\label{lupra3.8}
  \widehat{R} = 1 \otimes + \frac{1}{\kappa} r + {\cal O }
  (\frac{1}{\kappa^2})\, .
\end{equation}
If the classical r-matrix satisfies classical YB equation, the
quantum coproduct and quantum r-matrix can be described by a
twist function $T_\kappa$ in the following way:
\begin{eqnarray}\label{lupra3.9}
\Delta_\kappa (I_A) & = & T^{-1}_\kappa \circ (I_A \otimes 1 + 1
\otimes I_A) \circ T _\kappa \, ,
\\ \cr
R_\kappa & = & T^{-\kappa} \otimes (\tau T_\kappa) \, .
\end{eqnarray}
and one obtains coassociative Hopf algebra. In general case when
$r$ satisfies modified Yang-Baxter equation  the twist
(\ref{lupra3.9}) can be also performed \cite{drinf}, but the
coproduct is only quasicoassociative, and the formula
(\ref{lupra3.9}) describes a coproduct of quasi-Hopf quantum group
\cite{drinf}, with twist $T_\kappa$ called Drinfeld twist.

The problem of explicite quantizaton of classical Poincar\'{e}
r-matrices, listed in \cite{zak}, in particular the ones belonging
to the class (\ref{lupra3.1}), is not solved. Because in the case
of $\kappa$-deformed Poincar\'{e} algebra the Hopf-algebraic
structure is well known, mainly this quantum algebra is used to
describe the explicite formulae for deformed relativistic
symmetries. In particular it has been recently promoted a version
of doubly special relativistic theories \cite{bak}, \cite{jkg}
with the coalgebra structure determined by $\kappa$-deformed
Poincar\'{e} algebra.

\section{Discussion}
\setcounter{equation}{0}

 In this lecture we presented the variety of deformed addition laws
of fourmomentum based on Hopf--algebra structure. It should be
pointed out that recently due to some physical
arguments\footnote{The oldest one is related with validity of
$\kappa$-deformed mass-shall condition for large fourmomenta
describing macroscopic bodies, firstly pointed out to the authors
by I. Bia{\l}ynicki-Birula in 1997.} it was conjectured
\cite{ms2}, \cite{jkg2} that the addition law describing the
eigenstates of symmetry generators for composite systems should
not be described by a coproduct of Hopf-algebraic symmetry 
scheme.
In particular it has been argued in the framework of doubly
special relativity theory that the idea of having maximal momentum
or energy invariant under generalized Lorentz transformations can
not be sustained consistently with Hopf-algebraic coproduct rule.

In principle there are possible two approaches to the problem:

 i) the satisfactory solution should not violate Hopf algebra
structure. Quantum groups as Hopf algebras represent the 
extension
of the notion of symmetry to the case of symmetry transformations
with noncommutative symmetry parameter and describe the same
symmetries in arbitrary multiparticle sector. In such a case we
repeat the universal status of Poincar\'{e} group valid for any
multicomponent system, but we face some problems related with
physical interpretation.

ii) It should be however recalled that also without deformation
there were problems in description of interacting relativistic
multiparticle systems with exact classical Poincar\'{e} invariance
(see e.g. \cite{cjs}, \cite{hl}). In second approach one can
introduce therefore weaker nonuniversal space-time symmetries. In
such a case the modification of classical relativistic symmetries
needs a new algebraic concept which will reconcile e.g. physics at
macro-distances  with quantum -- deformed Planck lenght scale
physics, without complete discarding the structure of quantum
group \cite{smaj}. Such a possibility is provided by the
representation-dependent form of coalgebra,  depending e.g. on the
number of particles in the tensor product representation (see
\cite{ms2}) or in general on the labels of considered
representation. One can obtain such a framework by postulating the
deformation parameter $\kappa$ depends on ($\kappa$-deformed)
Casimir operators $\hat{C}_i$
\begin{equation}\label{lupra4.1}
\kappa\Longrightarrow \kappa(\hat{C}_i)\, .
\end{equation}
In particular representation $|{j_k}>$, where
\begin{equation}\label{lupra4.2}
\hat{C}_i|{j_k}> = j_i |{j_k}>\, .
\end{equation}
the operator-valued deformation parameter $\kappa(\hat{C}_i)$ is
replaced by its eigenvalue $\kappa(j_i)$.

The problem which should be considered further is the physical
choice the form of the function $\kappa(j_i)$. In particular if
the collection of labels ${j_i}$ describes quantum states which
converge to classical state, the function $\kappa(j_i)$ should
diverge to infinity, i.e. to the classical relativistic
no-deformation limit.

\end{document}